# Cost-Effective Quasi-Parallel Sensing Instrumentation for Industrial Chemical Species Tomography

Godwin Enemali, *Member, IEEE*, Rui Zhang, Hugh McCann, Chang Liu*, *Member, IEEE*

***Abstract*—Chemical Species Tomography (CST) has been widely applied for imaging of critical gas-phase parameters in industrial processes. To acquire high-fidelity images, CST is typically implemented by line-of-sight Wavelength Modulation Spectroscopy (WMS) measurements from multiple laser beams. The modulated transmission signal on each laser beam needs to be a) digitised by a high-speed analogue-to-digital converter (ADC); b) demodulated by a digital lock-in (DLI) module; and c) transferred to high-level processor for image reconstruction. Although a fully parallel data acquisition (DAQ) and signal processing system can achieve these functionalities with maximised temporal response, it leads to a highly complex, expensive and power-consuming instrumentation system with high potential for inconsistency between the sampled beams due to the electronics alone. In addition, the huge amount of spectral data sampled in parallel significantly burdens the communication process in industrial applications where *in situ* signal digitisation is distanced from the high-level data processing. To address these issues, a quasi-parallel sensing technique and electronic circuits were developed for industrial CST, in which the digitisation and demodulation of the multi-beam transmission signals are multiplexed over the high-frequency modulation within a wavelength scan. Our development not only maintains the temporal response of the fully parallel sensing scheme, but also facilitates the cost-effective implementation of industrial CST with very low complexity and reduced load on data transfer. The proposed technique is analytically proven, numerically examined by noise-contaminated CST simulations, and experimentally validated using a lab-scale CST system with 32 laser beams.**

## I. INTRODUCTION

INDUSTRIAL Process Tomography (IPT) has been widely applied to image, qualitatively and/or quantitatively, the behaviour of flows within processes in a non-intrusive manner.

Manuscript received Month xx, 2xxx; revised Month xx, xxxx; accepted Month x, xxxx. This work was supported in part by the UK Engineering and Physical Sciences Research Council (Platform Grant EP/P001661/1), and from the European Union (H2020 contract JTI-CS2-2017-CFP06-ENG-03-16). (Corresponding author: Chang Liu)

Various sensing modalities have been used, e.g. electrical tomography [1, 2], electromagnetic tomography [3], ultrasonic tomography [4, 5], x/γ-ray tomography [6], optical tomography [7], etc. Among these modalities, optical tomography utilising chemical species-dependent photon absorption is named Chemical Species Tomography (CST). Tunable diode laser absorption spectroscopy (TDLAS) is used in CST in a manner analogous to x-ray tomography, with the difference that, for incident light at an appropriately selected wavelength, the absorption measurements enable the reconstruction of the unknown spatial distribution of concentration of the target molecule [8]. Facilitated by the rapid technology development of the telecommunications industry, the last decade has witnessed accelerating advancement in the application of CST to solve industrial process engineering problems, e.g. vapour fuel imaging in internal combustion engines [9, 10], gas turbine exhaust imaging [11, 12], and power-plant boiler pollutant diagnosis [13].

Industrial CST is commonly implemented in harsh environments, suffering from significant noise in the line-of-sight (LoS) TDLAS measurements [14-16], dozens of which are required in order to achieve adequate spatial resolution [17]. The major sources of noise include electronic noise, optical noise, and environmental noise [18, 19]. To achieve better noise rejection capability, Calibration-Free Wavelength Modulation Spectroscopy (CF-WMS) is extensively adopted for implementing the LoS-TDLAS measurements in CST [20-24]. As shown in Fig. 1 (a), a high-frequency modulation (typically 40 kHz − 250 kHz) is added to a low-frequency wavelength scan (typically 1 Hz − 10 kHz). The CF-WMS scheme raises the signal detection band to a designated frequency range and then extracts the *n*-th harmonic signal by a demodulation process. Finally, path integrated absorbances for all the laser beams are calculated from the first harmonic normalised *n*-th harmonic signal, i.e. WMS-*nf*/1*f*. In most cases, WMS-2*f*/1*f* is favoured due to its stronger intensity.

The hardware implementation of a data acquisition (DAQ) system for CF-WMS requires three basic elements: (a) high-speed digitisation by an analogue-to-digital converter (ADC)

G. Enemali, R. Zhang, H. McCann and C. Liu are with the School of Engineering, University of Edinburgh, Edinburgh, EH9 3JL, U.K. (e-mail: C.Liu@ed.ac.uk)

with adequate bandwidth for sampling the WMS-2*f*, (b) digital lock-in (DLI) module for simultaneous demodulation of WMS-2*f* and WMS-1*f*, and (c) real-time transferring of the sampled WMS-2*f* and WMS-1*f* to high-level processors for post-processing and image reconstruction. To date, there are two typical sensing schemes to implement the CF-WMS method on all the laser beams for CST, i.e. fully-parallel (FP) [11, 25] and time-division-multiplexing (TDM) [15, 26] sensing schemes.

As shown in Fig. 1 (b), the FP sensing scheme detects simultaneously the laser transmission signals for all the laser beams by introducing an ADC and a DLI module for each beam. Although the FP sensing scheme can maximise the temporal resolution of the CST, its hardware implementation will inevitably lead to a highly complex, power-consuming and expensive instrumentation system. Considering a CST system, typically, with more than 30 laser beams, the large number of signal amplifiers and ADCs inevitably cause some inconsistency between the sampled beams. In addition, the huge amount of spectral data sampled in parallel significantly burdens the data transfer in industrial applications [11], in which *in situ* signal digitisation is generally located remotely from the high-level processors.

Alternatively, the TDM sensing scheme shown in Fig. 1 (c) simplifies the hardware by multiplexing the laser beams on the intervals of the neighbouring wavelength scans. The selected laser beam with its transmission signal is digitised by the ADC and then demodulated. However, the temporal resolution of such a CST system is degraded by the sequential sampling. In the case of turbulent flow imaging, the TDM scheme will potentially lead to insufficient characterisation of the instantaneous chemical reactions and heat transfer.

To address the above-mentioned challenges, we present here a quasi-parallel (QP) sensing technique and electronic circuits for industrial CST. The QP scheme is achieved by multiplexing the multi-beam signals over the high-frequency modulation within a wavelength scan, enabling the digitisation and demodulation of these beams with a single ADC and a single DLI module. The QP sensing scheme, initially addressed in [27], not only maintains the temporal response of the FP scheme, but also facilitates the cost-effective implementation of industrial CST with very low complexity and reduced load on data transfer. In this paper, we first illustrate the concept of the proposed scheme analytically. Then, numerical simulations are given, to examine the performance of the proposed scheme with noise-contaminated CST measurements. With electronic circuits designed to utilise a commercial off-the-shelf Field-Programmable Gate Array (FPGA) platform, the QP sensing scheme was experimentally validated using a lab-scale CST system with 32 laser beams.

## II. METHODOLOGY

### A. Background of WMS

The fundamentals of CF-WMS have been well developed for noise-resistant gas measurement [21]. To facilitate our presentation of the proposed WMS-based QP sensing scheme, the critical equations in terms of laser intensity modulation and frequency modulation in CF-WMS are reviewed. The time-varying diode laser output intensity $I_0(t)$ can be expressed as

$$I_0(t) = I_{0,s}(t) + I_{0,m}(t) \tag{1}$$

where the components $I_{0,s}(t)$ and $I_{0,m}(t)$ reflect, respectively, the injection current scan and current modulation at frequencies $f_s$ and $f_m$, given by Eqns. (2) and (3) where the same subscripts, $s$ and $m$, are used for the corresponding scan and modulation currents, $i_s$ and $i_m$.

$$I_{0,s}(t) = \bar{I}_0 \left[ \frac{1}{2} + i_{1,s} cos(2\pi f_s t + \phi_{1,s}) + i_{2,s} cos(4\pi f_s t + \phi_{2,s}) \right] \tag{2}$$

$$I_{0,m}(t) = \bar{I}_0 \left[ \frac{1}{2} + i_{1,m} cos(2\pi f_m t + \phi_{1,m}) + i_{2,m} cos(4\pi f_m t + \phi_{2,m}) \right] \tag{3}$$

where $\bar{I}_0$ is the average laser intensity, $i_1$ and $i_2$ are the amplitudes of the first- and second-order laser intensity modulation, respectively, and $\phi_1$ and $\phi_2$ are the phase shifts between the intensity modulation and frequency modulation for linear and non-linear components, respectively.

Accompanied by the intensity modulation, the output optical frequency of the laser, $v(t)$, can be modelled by

$$v(t) = \bar{v} + v_s(t) + v_m(t) \tag{4}$$

where $\bar{v}$ is the centre frequency, and $v_s(t)$ and $v_m(t)$ are the scanned and modulated frequencies, respectively. Similarly to $I_{0,s}(t)$ and $I_{0,m}(t)$, we express $v_s(t)$ and $v_m(t)$ as

$$v_s(t) = a_{1,s} cos(2\pi f_s t + \psi_{1,s}) + a_{2,s} cos(4\pi f_s t + \psi_{2,s}) \tag{5}$$

$$v_m(t) = a_{1,m} cos(2\pi f_m t + \psi_{1,m}) + a_{2,m} cos(4\pi f_m t + \psi_{2,m}) \tag{6}$$

where $a_1$ and $a_2$ are, respectively, first- and second-order amplitudes and again the subscripts $s$ and $m$ refer to scan and modulation, respectively. $\psi_1$ and $\psi_2$ are the first- and second-order phase shifts, respectively.

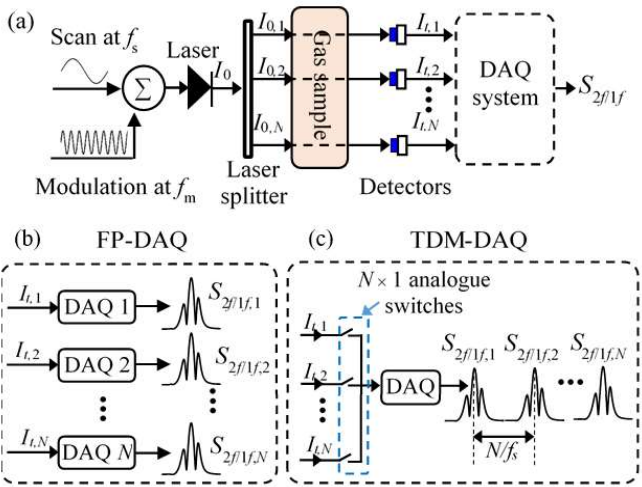

Figure 1: (a) Overview of a CST system with (b) Fully Parallel-DAQ sensing scheme, and (c) Time-Division Multiplexed-DAQ sensing scheme.

As shown in Fig. 1, the laser source is split into $N$ beams which pass through the target gas sample. Denoting the absorbance of the $i$-th laser beam by $\alpha_i(v(t))$ ($i = 1, 2, ..., N$), the relationship between its incident intensity $I_{0,i}(t)$ and transmitted intensity $I_{t,i}(t)$ follows Beer-Lambert's law and can be expressed by

$$I_{t,i}(t) = I_{0,i}(t)e^{-\alpha_i(v(t))} \tag{7}$$

Assuming, for the purpose of illustrating the basic sensing requirement, that the gas properties are uniform along each laser beam, $\alpha_i(v(t))$ can be calculated by

$$\alpha_i(v(t)) = L_i\phi(v)P_iS(T_i)X_i \tag{8}$$

where $L_i$ [cm] is the path length of the $i$-th laser beam through the subject, $\phi(v)$ [cm] is the line-shape function, $P_i$ [atm], $X_i$, and $T_i$ [K] are the pressure, molar fraction of the absorbing species, and temperature, respectively, and $S(T)$ [cm$^{-2}$atm$^{-1}$] is the line strength of the selected absorption line. In CST, it is typical that $X$ and $T$, and sometimes $P$, vary with position within the subject, and the aim of the technique is to recover the spatial distribution of $X$ and/or $T$.

### B. WMS-based QP Sensing Scheme for CST

As shown in Fig. 2, $I_{t,i}(t)$ from multiple beams are multiplexed within the wavelength scan at the resolution of the modulation period in the proposed QP sensing scheme. As a result, the multiplexed transmission signal, $I_{t,mux}(t)$ sampled by a single ADC and further processed by a DLI module can be expressed as

$$I_{t,mux}(t) = I_{t,i}(t)\big|_{t=(j-1)\times c/f_m}^{t=j\times c/f_m}, (j = 1, 2, ...) \tag{9}$$

where $j$ is the index of the multiplexing operation, and $c$ is the number of modulation periods sampled per beam. The index of laser beam being sampled, $i$, is calculated by

$$i = j - \lfloor(j-1)/N\rfloor \times N \tag{10}$$

where $N$ is the total number of beams multiplexed over a wavelength scan. The operand $\lfloor \bullet \rfloor$ is the floor function. As a result, $I_{t,mux}(t)$ contains the intermittent $I_{t,i}(t)$ from all the $N$ multiplexed laser beams within a wavelength scan. As the temporal resolution of a CST system depends on the frequency of the wavelength scan, the proposed QP sensing scheme can also maintain the maximum temporal resolution achieved using the FP sensing scheme, provided that all N beams are adequately sampled within a single wavelength scan. Fig. 2 explicitly illustrates the QP sensing scheme for $c = 1$ and $N = 4$.

To achieve the CF-WMS scheme, the WMS-1$f$ and WMS-2$f$ signals are extracted from the digitised $I_{t,mux}(t)$, denoted as $I_{t,mux}(d)$, using the DLI module. The in-phase and quadrature components of the WMS-1$f$, $X_{1f,mux}$ and $Y_{1f,mux}$, and WMS-2$f$, $X_{2f,mux}$ and $Y_{2f,mux}$, can be obtained by [28]

$$X_{1f,mux} = \sum_{d=0}^{D-1} I_{t,mux}[d] \times R_{I,1f}[d] \tag{11}$$

$$Y_{1f,mux} = \sum_{d=0}^{D-1} I_{t,mux}[d] \times R_{Q,1f}[d] \tag{12}$$

$$X_{2f,mux} = \sum_{d=0}^{D-1} I_{t,mux}[d] \times R_{I,2f}[d] \tag{13}$$

$$Y_{2f,mux} = \sum_{d=0}^{D-1} I_{t,mux}[d] \times R_{Q,2f}[d] \tag{14}$$

where $R_{I,1f}[d]$ and $R_{Q,1f}[d]$ ($R_{I,2f}[d]$ and $R_{Q,2f}[d]$) are sinusoidal reference signals with $\pi/2$ difference in phase at $f_m$ ($2f_m$), respectively. The total number of digitised samples $D = c \times \frac{f_d}{f_m}$, where $f_d$ is the sampling frequency.

Similar to the demodulation of $I_{t,mux}(t)$, the non-absorbing background signal $I_{0,mux}(t)$ is sampled and demodulated to obtain the absorption-free in-phase and quadrature components, $X_{1f,mux}^0$, $Y_{1f,mux}^0$, $X_{2f,mux}^0$ and $Y_{2f,mux}^0$, respectively. The multiplexed WMS-2$f$/1$f$ signal, $^{mux}S_{2f/1f}$, is calculated as

$$^{mux}S_{2f/1f} = \sqrt{\left(\frac{X_{2f,mux}}{R_{1f,mux}} - \frac{X_{2f,mux}^0}{R_{1f,mux}^0}\right)^2 + \left(\frac{X_{2f,mux}}{R_{1f,mux}} - \frac{X_{2f,mux}^0}{R_{1f,mux}^0}\right)^2} \tag{15}$$

where

$$R_{1f,mux}^0 = \sqrt{\left(X_{1f,mux}^0\right)^2 + \left(Y_{1f,mux}^0\right)^2} \tag{16}$$

$$R_{1f,mux} = \sqrt{\left(X_{1f,mux}\right)^2 + \left(Y_{1f,mux}\right)^2} \tag{17}$$

Each sample of $^{mux}S_{2f/1f}$ shown in Fig. 2 is demodulated from the accumulated $c$ modulation periods. As a result, the samples of $^{mux}S_{2f/1f}$ from the multiplexed signal are obtained in sequence. For example, $^{mux}S_{2f/1f}$ obtained in Fig. 2 contains the samples from the 4 multiplexed beams and can be expressed as

$$^{mux}S_{2f/1f} = \{... \beta_1, \gamma_1, \delta_1, \varepsilon_1, \beta_2, \gamma_2, \delta_2, \varepsilon_2, ...\}. \tag{18}$$

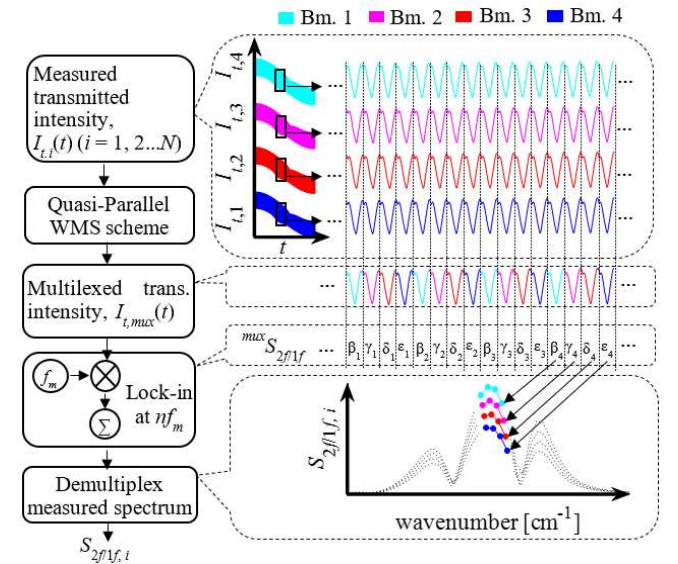

Figure 2: Flowchart of the QP-WMS sensing scheme demonstrated for four-beam multiplexing.

where $\beta_j$, $\gamma_j$, $\delta_j$, $\varepsilon_j$, denotes the samples of $^{mux}S_{2f/1f}$ obtained from the 4 multiplexed laser beams with the $j$-th multiplexing operation, respectively.

Then, the $^{mux}S_{2f/1f}$ is demultiplexed to recover the $S_{2f/1f}$ spectrum for each beam, $S_{2f/1f,i}$ $(i = 1,2,3,4)$, as

$$S_{2f/1f,1} = \{\dots\ \beta_1, \beta_2, \dots\} \tag{19}$$

$$S_{2f/1f,2} = \{\dots\ \gamma_1, \gamma_2, \dots\} \tag{20}$$

$$S_{2f/1f,3} = \{\dots\ \delta_1, \delta_2, \dots\} \tag{21}$$

$$S_{2f/1f,4} = \{\dots\ \varepsilon_1, \varepsilon_2, \dots\} \tag{22}$$

## III. NUMERICAL SIMULATION

### A. Simulation Set up

Numerical simulation was carried out to validate the proposed QP sensing scheme using the water absorption transition at 7185.6 cm$^{-1}$. Four laser beams were used in the simulation. Along the laser path with length of 36 cm, temperature and pressure were assumed to be uniform at 293 K and 1 atm, respectively. The water concentrations on the four laser paths were assumed to be 0.8% (mole fraction), 0.7%, 0.6% and 0.5%, respectively. $f_s$ and $f_m$ were set to 31.25 Hz and 62.5 kHz respectively, i.e. $f_m$ was 2,000 times $f_s$, ensuring that all beams could be sampled many times during the wavelength scan. $a_{1,s}, a_{1,m}, \bar{I}_0, i_{1,s}, i_{1,m}$ were set as 0.25, 0.006, 0.5, 0.2 and 0.1 respectively. As the absorption feature is scanned twice within each sinusoidal scan, only the falling part of the scan is used in this work, as shown in Fig. 3. According to Eqn. (9), the transmission signals, $I_{t,i}(t)$ $(i = 1,2,3,4)$, were multiplexed to obtain $I_{t,mux}(t)$. By following the similar procedures, $I_{0,mux}(t)$ was also obtained. Both $I_{t,mux}(t)$ and $I_{0,mux}(t)$ were digitised at a rate of $f_d = 15.625$ Mega Samples per second (MSps) with a single ADC and subsequently demodulated by a DLI module with $c = 2$ to obtain $^{mux}S_{2f/1f}$.

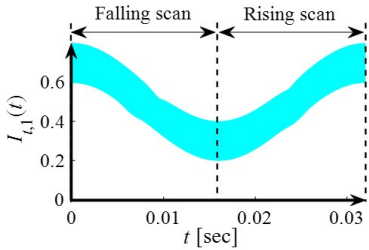

Figure 3: Simulated raw transmitted signal for beam 1.

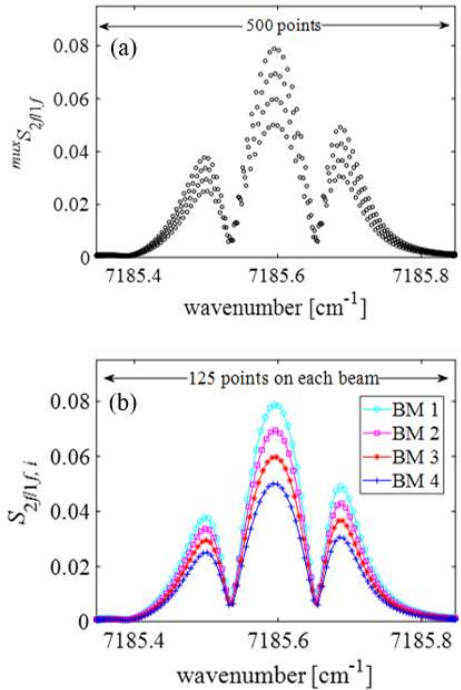

Figure 4. For the four-beam simulated sensing using QP-WMS, (a) the demodulated $^{mux}S_{2f/1f}$ and (b) the demultiplexed $S_{2f/1f,i}$.

As shown in Fig. 4 (a), the demodulated $^{mux}S_{2f/1f}$ contains 500 samples on the wavelength. Using Eqns. (18) to (22), $S_{2f/1f,i}$ $(i = 1,2,3,4)$, each with 125 samples shown in Fig. 4 (b), is obtained after the demultiplexing. The harmonic information for the four beams can be obtained using a single ADC-DLI pair with a sampling interval of 32 $\mu$s ($4/f_m/c$) between the neighbouring beams. Therefore, the temporal response of the QP sensing scheme is much superior to the TDM scheme that samples the neighbouring beams with a sampling interval of 32 ms ($1/f_s$) with the same settings. In comparison with the FP sensing scheme, the proposed QP scheme saves 75% hardware resources in terms of both the ADC-DLI pairs and memory space, whilst maintaining the frame rate of the CST system, i.e. $f_s$. The reduced data throughput also alleviates the requirement placed on bandwidth, thus facilitating remote data transfer in real time.

### B. Numerical Results and Discussions

The performance of the proposed QP sensing scheme is examined by considering the practical measurements, in which the transmission for each laser beam is contaminated by various forms of noise. The sources of noise can be characterised by frequency-dependent noise and frequency-independent noise, respectively. The frequency-dependent noise mainly contains pink noise ($1/f$ noise), e.g. laser and detector excess noise, and low-frequency environmental noise generally caused by background scattering and flow-induced beam steering. The frequency-independent noise, which is known to have a white noise spectrum, is dominated by the thermal noise from the detector, signal amplification and digitisation circuits [18]. In the simulation, the Signal-to-Noise Ratio (SNR) for the environmental noise was set to 15 dB lower, with respect to $I_t$. As shown in Fig. 5, this noise level was quantified by measuring the laser intensity fluctuation when a turbulent heated flow was imposed on a laser beam with stable intensity, using the ADC of the Red Pitaya with a bandwidth of 40 MHz. In addition, the combined pink and white noise was set to 56 dB lower with respect to $I_t$ in the simulation. As shown in Table I, being calculated at the frequency of 62.5 kHz, consistent with the modulation frequency, this noise level can be justified by the noise performance of the major components in the detection system that will be detailed in Section IV.

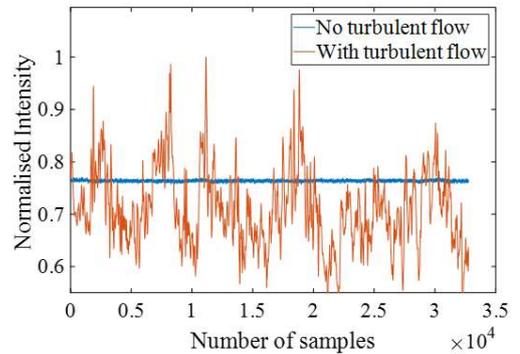

Figure 5: Quantification of laser intensity fluctuations when a turbulent heated flow was imposed on a laser beam with stable intensity.

TABLE I
NOISE PERFORMANCE OF MAJOR COMPONENTS

| Photo-detector (G12182-003K, Hamamatsu) | Pre-amp circuit (AD8033) | PGA (THS7002, Texas Instruments) | Red Pitaya [30] |
|---|---|---|---|
| 6.5e-13 W/√Hz | 11 nV/√Hz | 1.7 nV/√Hz | 71.3 dB (SNR) |

Fig. 6 shows a flowchart of the simulated implementation of the QP sensing scheme with sequentially imposed multiple sources of noise. As the low-frequency environmental noise is introduced during beam propagation, it is firstly added to the noise-free transmission for the $i$-th beam. Subsequently, the pink noise and white noise, mainly introduced by the detector and data acquisition (DAQ), are added, resulting in the noise-contaminated transmission, $I_{t,i}(t)$. Finally, the multi-beam transmission signals are multiplexed and demodulated to obtain the $S_{2f/1f,i}$, from which the path integrated absorbances can be obtained for image reconstruction [29].

To be specific, Fig. 7 (a-d) demonstrates the noise-contaminated transmission signals for the four beams noted above in Section III A. All the three sources of noise were randomly generated within Gaussian distributions with zero mean and standard deviation set to the fixed noise levels, i.e. 15 dB for environmental noise, and 56 dB for pink and white noise. To validate the proposed scheme, $I_{t,i}(t)$ ($i = 1, 2, 3, 4$) were demodulated and fitted by using both the QP and FP sensing schemes. As shown in Figs. 7 (e-h) and (i-l), the proposed QP scheme, with only one quarter of the total number of samples in comparison with the FP sensing scheme, yields fitting results with similar residuals. To be quantitative and statistical, Table II shows the mean and standard deviation (std.) of the residual difference between the demodulated and fitted $S_{2f/1f,i}$ data for both the FP and QP sensing schemes using 500 repetitive simulations. Both the mean and std. values are very similar for both schemes with a maximum difference of 0.56% and 1.91%, respectively.

TABLE II
COMPARISON OF FITTING RESIDUALS BETWEEN FP AND QP SCHEME

| Beam | Parameter | FP | QP | Difference (% |
|---|---|---|---|---|
| BM 1 | Mean ($10^{-3}$) | 6.2017 | 6.1988 | 0.0468 |
| | Std. ($10^{-4}$) | 2.3082 | 2.3046 | 0.1560 |
| BM 2 | Mean ($10^{-3}$) | 5.5534 | 5.5310 | 0.4034 |
| | Std. ($10^{-4}$) | 3.5778 | 3.5157 | 1.7357 |
| BM 3 | Mean ($10^{-3}$) | 4.8518 | 4.8280 | 0.4905 |
| | Std. ($10^{-4}$) | 3.4364 | 3.4686 | 0.9370 |
| BM 4 | Mean ($10^{-3}$) | 4.1496 | 4.1264 | 0.5591 |
| | Std. ($10^{-4}$) | 3.5033 | 3.4365 | 1.9068 |

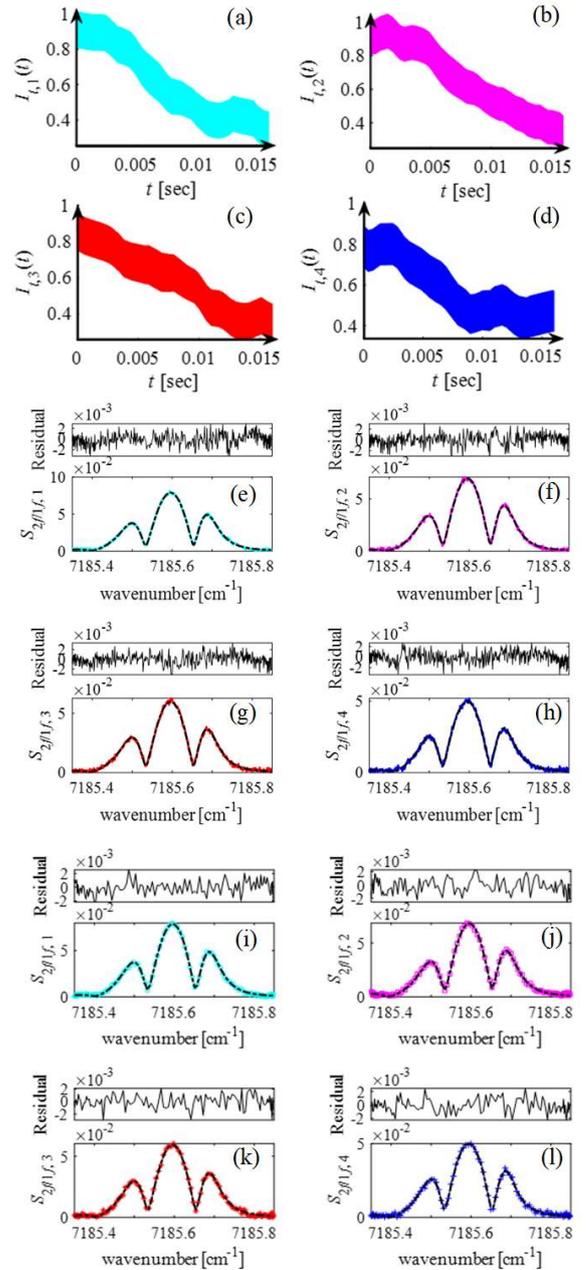

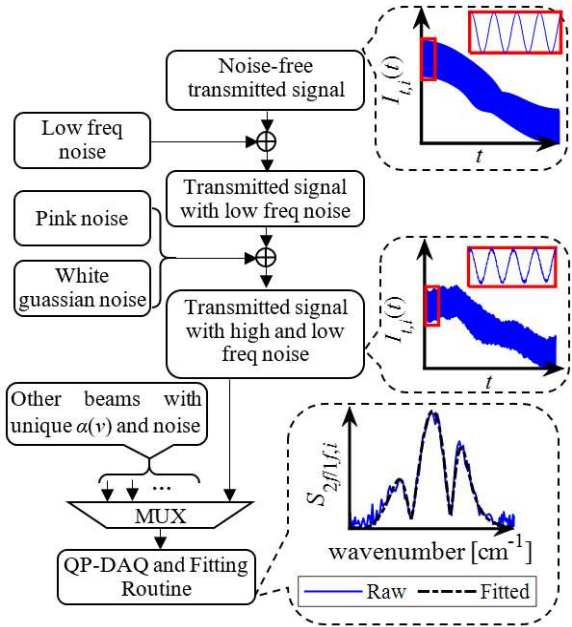

Figure 6: Flowchart of the simulated implementation of the QP sensing scheme with sequentially introduced multiple sources of noise.

Figure 7. Implementation of both the FP and QP sensing schemes on the same laser transmission signals. (a-d) shows the simulated noise-contaminated transmission signals for the four laser beams. (e-h) and (i-l) shows the demodulated and fitted $S_{2f/1f,i}$ using the FP and QP sensing schemes, respectively.

## IV. HARDWARE AND FIRMWARE IMPLEMENTATION

For hardware implementation of the QP sensing scheme, we designed a signal conditioning circuit (SCC), which consists of (a) a 4-to-1 multiplexer (ADG704, Analogue Devices) to produce the multiplexed transmission signal, $I_{t,mux}(t)$, from the four laser beams, and (b) a programmable gain amplifier (PGA) (THS7002, Texas Instruments) that further conditions the multiplexed signal. As shown in Fig. 8 (a), the control bits for the SCC are derived from the digital inputs/outputs (DIOs) on a commercial off-the-shelf FPGA platform, the Red Pitaya (RP) [30]. The analogue output from the SCC is digitised by a 14-bit 125 MSps ADC that is integral to the RP.

The worst-case response time, $t_{mux}$, of the ADG704 is 33 ns. To guarantee signal integrity, $t_{mux}$ must satisfy

$$t_{mux} < 1/f_d. \quad (23)$$

That is to say, the multiplexed signal must be stabilised between the neighbouring samples digitised by the ADC. As a result, the theoretical maximum sampling frequency $f_d$ supported in the design is 30.3 MHz. Its switching control is implemented by a multiplexer switching timing control (MSTC) logic. The MSTC synchronises the switching operation of the multiplexer to the ADC by taking $c$, $f_d$, and $f_m$ into account, generating precisely timed logic signals that select the laser beam to be sampled per modulation period.

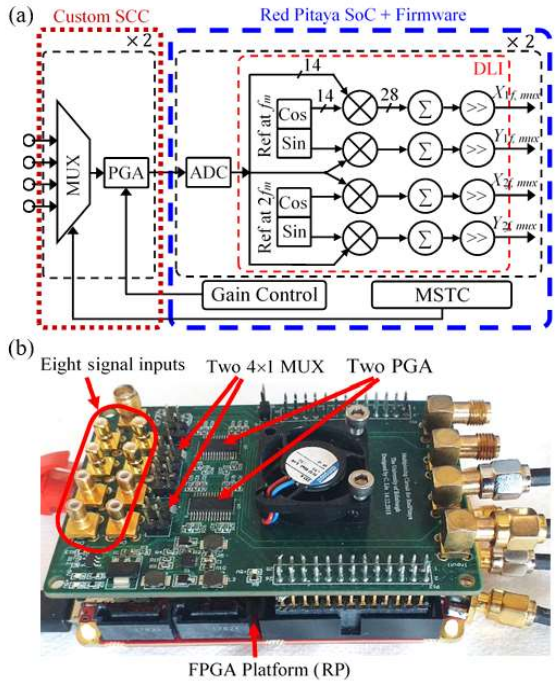

Figure 8: Hardware implementation of QP sensing scheme: (a) schematics of the SCC and the FPGA platform; (b) photograph of the hardware circuits, indicating the layout.

Subsequently, the digitised $I_{t,mux}(t)$ is demodulated by the DLI customised on the FPGA (Xilinx 7 series ZYNQ 7010) of the RP at the frequencies of $1f_m$ and $2f_m$. For each frequency, two 14-bit sinusoidal reference signals with phase difference of 90 degrees are generated using Xilinx's DDS IPs and then multiplied by the 14-bit digitised $I_{t,mux}(t)$. According to Eqns. (11)-(14), the quadrature components, $X_{1f,mux}$, $Y_{1f,mux}$, $X_{2f,mux}$ and $Y_{2f,mux}$, can be obtained by accumulating the 28-bit multiplied signals for $c$ modulation periods. The accumulation process increases each quadrature component to $[28 + log(c * f_s/f_m)]$ bits. As a standard 32-bit embedded processor is used for data post-processing and transfer, shift operation, as shown in Fig. 8 (a), is adopted to discard excess less significant bits. As a result, overflow can be avoided while maintaining minimal loss in precision. Finally, the quadrature components are transferred to the PC via Ethernet protocol.

As each RP is equipped with two simultaneous ADCs, the developed SCC and the DLI firmware are also duplicated, enabling two sets of 4-beam QP sensing. Fig. 8 (b) shows the hardware circuits, which enable a very compact implementation. The FPGA resources incurred by the QP sensing scheme in addition to the base firmware of RP are 1334 Flip Flops, 1456 Look-up Tables, 19 Block RAMs and 8 DSPs.

## V. EXPERIMENTS

### A. Experimental Set Up

To experimentally validate the proposed QP sensing system, a CST system with 32 laser beams was developed to image the two-dimensional (2D) distribution of water vapour ($H_2O$) concentration. As each measurement hub (Fig. 9(a)) is capable of measuring 8 beams with the QP sensing scheme, as discussed above and shown in Fig. 8, four hubs were utilised. A distributed feedback laser diode (NLK1E5GAAA, NTT Electronics) working at 7185.6 cm$^{-1}$ was used for $H_2O$ sensing in the experiment reported here. The laser was fed with a signal consisting of a scan sinusoid at $f_s = 31.25$ Hz and modulation sinusoid at $f_m = 62.5$ kHz from an arbitrary signal generator (ASG). Using a 32-way fibre splitter, the output laser from the pigtailed laser diode was equally split into 32 channels, each collimated by a fibre collimator (CFC5-C, Thorlabs) for launch across the subject. As shown in Fig. 9 (a), the 32 laser beams were arranged in 4 equiangular projections (45°) with 8 parallel beams in each projection. The neighbouring beam spacing, $d$, is 1.8 cm, while the distance $D$ between the emitter and detector of each beam is 36.76 cm. Each transmitted laser beam was incident on a dedicated photodetector (G12182-003K, Hamamatsu). The 32 transmission signals were processed in QP mode, with a dedicated hub for each group of 8 signals, multiplexing 4 signals to each ADC, sampling at 15.625 MSps. With $c = 2$, a total of 500 samples were accumulated for each DLI operation to obtain 1000 wavelength samples in $^{mux}S_{2f/1f}$ per frame on each hub. Finally, signals were demultiplexed to obtain $S_{2f/1f,i}$ for all 32 beams, leading to 125 wavelength points in falling and rising part of the scan, respectively. The data acquisition was synchronised to the laser driver by means of a trigger generated by the ASG at $f_s$. Data transfer to a remote PC from the 4 hubs was achieved using an Ethernet switch (GO-SW-5G, D-Link), enabling transfer of $4000 \times 32$ bits per hub per frame. The tomographic frame rate was 31.25 fps.

### B. Experimental Results and Discussion

To examine the noise performance of the developed electronics, the $S_{2f/1f,i}$ ($i = 1, 2, 3, 4$) were measured using a single DAQ channel for 100 times in room air without any artificially generated $H_2O$ absorption. In this case, there is no flow-induced uncertainty imposed on the laser beams. As shown in Fig. 10, each $S_{2f/1f,i}$ contains 125 wavelength samples with the proposed QP sensing scheme. Fig. 10 shows the mean value of each $S_{2f/1f,i}$, while the error bars show the standard deviation. As the temperature, $H_2O$ concentration and path length are same for the four laser beams, the peak values of the sampled $S_{2f/1f,i}$ ($i = 1, 2, ..., 4$) are very close, with a maximum difference of 2.3%. The maximum standard deviation of $S_{2f/1f,i}$ observed for any of the four beams is 0.0052, demonstrating excellent noise suppression by the developed QP sensing instrumentation.

A phantom (i.e. known) 2D $H_2O$ vapour distribution was generated using a container filled with hot water, and the measurement plane placed vertically above it, in a simple proof-of-concept experiment. As shown in Fig. 9 (b), the container was placed at the centre of the Region of Interest (RoI). Fig. 11 shows the $S_{2f/1f,i}$ sampled from the same four laser beams, extracted from a single frame of data. Stronger absorption can be observed for beams 3 and 4, due to their penetration of the evaporated $H_2O$ with longer path length. To reconstruct the 2D distribution of $H_2O$ concentration, the path integrated absorbances from the 32 laser beams were calculated by extracting the peak values of $S_{2f/1f,i}$ [25, 29]

$$A_{m,i} = P_{m,i}\left(\frac{A_s}{P_s}\right) \tag{24}$$

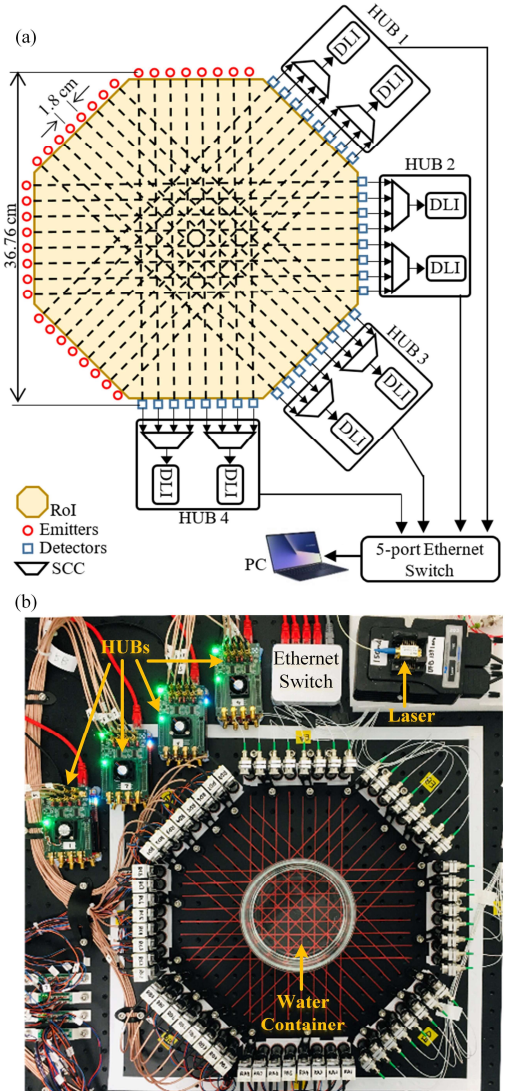

Figure 9: Hardware implementation of QP sensing scheme. (a) shows the schematic of SCC and the FPGA platform. (b) shows the picture of the hardware circuits.

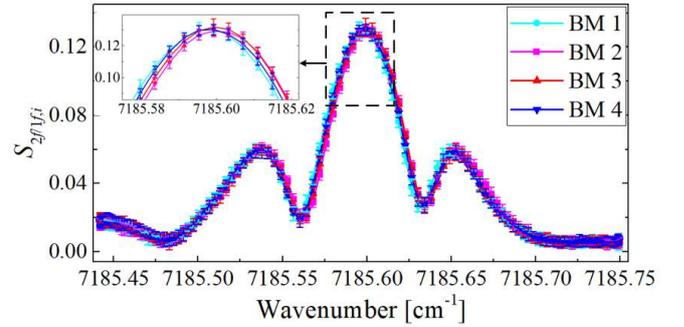

Figure 10: Repetitively measured $S_{2f/1f,i}$ ($i = 1, 2, 3, 4$) for 100 times using the developed QP sensing scheme and instrumentation on a single DAQ channel.

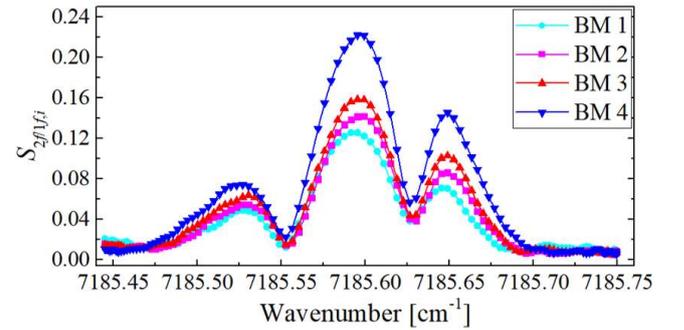

Figure 11: Measured $S_{2f/1f,i}$ ($i = 1, 2, 3, 4$) for reconstructing a single frame of image using the developed QP sensing scheme and instrumentation on a single DAQ channel.

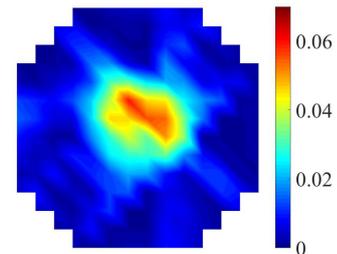

Fig. 12 Reconstructed 2D distribution of $H_2O$ concentration using the developed QP sensing scheme and instrumentation.

where $A_{m,i}$ and $P_{m,i}$ are the path integrated absorbance and the peak value of $S_{2f/1f,i}$ for the $i$-th laser beam, respectively. $A_s$ and $P_s$ are the simulated path integrated absorbance and the peak value of $S_{2f/1f}$ calculated using the HITRAN 2016 database, respectively [31].

Fig. 12 shows the resulting image of $H_2O$ concentration reconstructed using the Simultaneous Algebraic Reconstruction Technique (SART) with detailed procedures illustrated in our previous research [32]. The reconstructed image exhibits good quality with few artefacts. In addition, the reconstructed image agrees well with the phantom in terms of the location of the high $H_2O$ concentration region relative to room air, further indicating the effectiveness of the proposed QP sensing scheme and the developed instrumentation.

## VI. CONCLUSION

This paper introduces a cost-effective QP sensing technique and instrumentation scheme, developed for industrial application of CST. The new development implements, for the first time, multiplexing of multi-beam transmission signals over the high-frequency modulation within each wavelength scan, followed by digitisation and demodulation. Consequently, it significantly reduces the hardware/firmware complexity of CST implementation and its load on data transfer. In particular, this scheme enables a large number of laser beams to be employed in industrial systems robustly and cost-effectively.

The proposed technique was analytically demonstrated and then numerically examined by simulations. In the simulation, electronic, optical and environmental noise were added to the noise-free transmission signal. The $S_{2f/1f}$ was obtained using both the QP and FP sensing schemes, and fitted for quantitative comparison. The results show the proposed QP scheme saves three quarters of the wavelength samples in comparison with the FP sensing scheme, but with excellent fitting results with maximum differences of 0.56 % and 1.91 % in the mean and standard deviation of the fitting residuals.

Furthermore, electronic circuits were developed and a lab-scale CST system with 32 laser beams was set up to experimentally validate the proposed scheme. Experimental results further demonstrated that high-fidelity harmonic signals can be obtained using the proposed QP sensing scheme. Finally, the path integrated absorbances extracted from the harmonic signals were used to reconstruct a 2D distribution of $H_2O$ concentration, demonstrating the effectiveness of the developed technique and instrument.